\documentclass{ws-ijmpe}

\begin{document}
\newcommand{\pom}{\tt I\! P}
\newcommand{\beq}{\begin{equation}}
\newcommand{\eeq}{\end{equation}}

\markboth{M.B. Gay Ducati, M.M. Machado and M.V.T. Machado}{Investigating Diffractive W Production In Hadron-Hadron Colliders At High Energies}

\catchline{}{}{}{}{}

\title{{\bf INVESTIGATING DIFFRACTIVE $W$ PRODUCTION IN\\HADRON-HADRON COLLISIONS  AT HIGH ENERGIES}}

\author{\footnotesize {\bf M.B. GAY DUCATI} and {\bf M.M. MACHADO}}

\address{High Energy Phenomenology Group, GFPAE
Instituto de F\'isica, \\ Universidade Federal do Rio Grande do Sul,Caixa Postal 91501, CEP 91501-970,\\ Porto Alegre -RS, Brazil 
beatriz.gay@if.ufrgs,br, melo.machado@ufrgs.br}

\author{{\bf M.V.T. MACHADO}}

\address{Centro de Ci\^encias Exatas e Tecnol\'ogicas, Universidade Federal do Pampa \\
Campus de Bag\'e, Rua Carlos Barbosa. CEP 96400-970. Bag\'e, RS, Brazil\\
mmachado.unipampa@ufpel.edu.br}

\maketitle

\begin{history}
\received{(25 May 2007)}
\accepted{(18 June 2007)}
\end{history}
\vspace{-0.2cm}
\begin{abstract}
We compute the hard diffractive
hadroproduction of bosons $W^{\pm}$ at high energies using Regge factorization and taking into account multiple Pomeron exchange corrections. The ratio of diffractive to non-diffractive $W$ production agrees with the current Tevatron data and a prediction for the LHC is presented.
\end{abstract}
\vspace{-0.5cm}
\section{Introduction}
Diffractive processes in hadron collisions are well described by Regge theory in terms of the exchange of a Pomeron with vacuum quantum numbers.\cite{Collins} A good channel attracting much attention is the use of hard scattering to resolve the quark and gluon content in the Pomeron.\cite{IS}  Observations of diffractive deep inelastic scattering at HERA have increased the knowledge about the QCD Pomeron, providing us with the diffractive distributions of singlet quarks and gluons (Dpdf's) in the Pomeron.\cite{H1diff} At high energies, there is strong suppression of the single-Pomeron Born cross section due to multi-Pomeron exchanges. At the Tevatron, the
suppression factor is about 0.05--0.2, whereas for the LHC the suppression is of order 0.08--0.1 (see, for instance \cite{KKMR} for reliable estimates).

In this contribution we summarize the results obtained in \cite{ggm_prd}, where the diffractive $W$ production is computed for the Tevatron energy and estimates are provided for the CERN LHC experiment.  We summarize the details concerning the dpdf's in the Pomeron, extracted recently in DESY-HERA, and present theoretical estimates for the gap survival probability factor. In the last section, we present the numerical results for the diffractive ratio in Tevatron and extrapolate them to the LHC energy.
\vspace{-0.4cm}
\section{Diffractive Hadroproduction of $W^{\pm}$}

For the hard diffractive processes we will consider the Ingelman-Schlein (IS) picture, \cite{IS} where
the Pomeron structure is probed. The generic cross section for a process in which partons of two hadrons, $A$ and $B$, interact to produce a massive electroweak boson, $ A + B \rightarrow W^{\pm} + X$, reads as
\begin{eqnarray}
\frac{d \sigma}{dx_a\,dx_b} = \sum_{a,b} f_{a/A}(x_a,
\mu^2)\, f_{b/B}(x_b, \mu^2)\, \frac{d\hat{\sigma}(ab\rightarrow W\,X)}{d\hat{t}}\,,
\label{gen}
\end{eqnarray}
where $x_i f_{i/h}(x_i, \mu^2)$ is the distribution function of a parton of flavour $i=a,b$ in the hadron $h=A,B$.  The quantity $d\hat{\sigma}/d\hat{t}$ gives the elementary hard cross section of the corresponding subprocess and $\mu^2=M_{W}^2$ is the hard scale in the QCD evolution. In the expression for
diffractive processes, one assumes that one of the hadrons, say
hadron $A$, emits a Pomeron whose partons interact with partons of the hadron $B$.
Thus the parton distribution  $x_a f_{a/A}(x_a, \mu^2)$ in
Eq.~(\ref{gen}) is replaced by the convolution between a
distribution of partons in the Pomeron, $\beta f_{a/{\pom}}(\beta,
\mu^2)$, and the ``emission rate" of Pomerons by the hadron, $f_{{\pom}/h}(x_{{\pom}},t)$. The last quantity, $f_{{\pom}/h}(x_{{\pom}},t)$, is the Pomeron flux factor and its explicit formulation is described in
terms of Regge theory. Therefore, we can rewrite the parton distribution as
\begin{eqnarray}
\label{convoP}
x_a f_{a/A}(x_a, \,\mu^2)\ =\ \int dx_{{\pom}} \
\bar{f}(x_{{\pom}})\, {\frac{x_a}{x_{{\pom}}}}\, f_{a/{\pom}}
({\frac{x_a}{x_{{\pom}}}}, \mu^2).
\end{eqnarray}
where we have defined the quantity $\bar{f} (x_{{\pom}}) \equiv \int_{-\infty}^0 dt\
f_{{\pom/A}}(x_{{\pom}},t)$.

Concerning the $W^{\pm}$ diffractive production, one considers the reaction
$p + {\bar p}(p) \rightarrow p + \ W (\rightarrow e\ \nu ) + \ X$, assuming that a Pomeron emitted by a proton in
the positive $z$ direction interacts with a $\bar p$ (or a $p$) producing $W^{\pm}$
that subsequently decays into $e^{\pm}\ \nu$. By using the same concept of the convoluted structure function, the
diffractive cross section for the inclusive lepton production becomes
\begin{eqnarray}
\frac{d\sigma^{\mathrm{SD}}_{\mathrm{lepton}}}{d\eta_e}= \sum_{a,b}
\int \frac{dx_{\pom}}{x_{\pom}}\, \bar{f}(x_{\pom})
\int dE_T \ f_{a/{\pom}}(x_a, \,\mu^2)\,f_{b/\bar{p}(p)}(x_b, \,\mu^2)\
\left[\frac{ V_{ab}^2\ G_F^2}{6\ s\ \Gamma_W}\right]\ \frac{\hat{t}^2}
{\sqrt{A^2-1}}
\label{dsw}\nonumber
\end{eqnarray}
where
\begin{eqnarray}
x_a = \frac{M_W\ e^{\eta_e}}{(\sqrt{s}\ x_{{\tt I\! P}})}\ \left[A \pm
\sqrt{(A^2-1)}\right]
\label{xaw},\,\,\,x_b = \frac{M_W\ e^{-\eta_e}}{\sqrt{s}}\ \left[A \mp \sqrt{(A^2-1)}\right],
\end{eqnarray}
with $A={M_W}/{2 E_T}$, $E_T$ being the lepton transverse energy, $G_F$ is the Fermi constant and the hard scale $\mu^2=M_W^2$. The quantity  $V_{ab}$ is the Cabibbo-Kobayashi-Maskawa matrix element and $\hat{t}=-E_T\ M_W\ \left[A+\sqrt{(A^2-1)}\right]$. The upper signs in Eqs.~(\ref{xaw}) refer to $W^+$ production (that is, $e^+$ detection). The corresponding
cross section for $W^-$ is obtained by using the lower signs and ${\hat t}
\leftrightarrow {\hat u}$.

An important element in the calculation is the Pomeron flux factor, introduced in
Eq.~(\ref{convoP}). We take the experimental analysis of the diffractive structure function, \cite{H1diff} $f_{\pom/p}(x_{\pom}, t) = A_{\pom}\frac{e^{B_{\pom} t}}{x_{\pom}^{2\alpha_{\pom} (t)-1}}$,
where  the Pomeron trajectory is assumed to be linear,
$\alpha_{\pom} (t)= \alpha_{\pom} (0) + \alpha_{\pom}^\prime t$, and the parameters
$B_{\pom}$ and $\alpha_{\pom}^\prime$ and their uncertainties are obtained from
fits to H1 FPS data. \cite{H1diff} In our estimates, we will consider the dpdf's recently obtained by the H1 Collaboration at DESY-HERA. \cite{H1diff} The Pomeron structure function has been modeled in terms of a
light flavour singlet distribution $\Sigma(z)$, consisting of $u$, $d$ and $s$
quarks and anti-quarks
with $u=d=s=\bar{u}=\bar{d}=\bar{s}$,
and a gluon distribution $g(z)$.  Here, $z$ is the longitudinal momentum
fraction of the parton entering the hard sub-process
with respect to the diffractive
exchange, such that $z=\beta$
for the lowest order quark-parton model process,
whereas $0<\beta<z$ for higher order processes.
The quark singlet and gluon distributions are parameterized
at $Q_0^2$ with the general form, $
z f_i (z,\,Q_0^2) = A_i \, z^{B_i} \, (1 - z)^{C_i} \,\exp\left[{- \frac{0.01}{(1-z)}}\right]$,
where the last exponential factor ensures that the diffractive pdf's vanish
at $z = 1$. The charm and beauty quarks are treated as massive, appearing through boson gluon fusion-type processes up to order $\alpha_s^2$.

It has been known for a long time that factorization does not  
necessarily hold for diffractive production processes. \cite{KKMR} The suppression  
of the single-Pomeron Born cross section due to the multi-Pomeron  
contributions depends, in general, on the particular hard process.  We  
will consider this suppression through a gap survival probability  
factor, $<\!|S|^2\!>$, using two theoretical estimates for this factor: (a) model of \cite{KKMR} (labeled KMR), which considers a two-channel eikonal model. The survival probability is computed for single, central and double diffractive processes at several energies. We will consider the results for single diffractive processes, where $<\!|S|^2\!>_{\mathrm{KMR}}=0.15$ for $\sqrt{s}=1.8$ TeV (Tevatron) and $<\!|S|^2\!>_{\mathrm{KMR}}=0.09$ for $\sqrt{s}=14$ TeV (LHC). (b) The second theoretical estimate is from \cite{GLM} (labeled GLM), which considers a single channel eikonal approach, where $<\!|S|^2\!>_{\mathrm{GLM}}=0.126$ for $\sqrt{s}=1.8$ TeV (Tevatron) and $<\!|S|^2\!>_{\mathrm{GLM}}=0.081$ for $\sqrt{s}=14$ TeV (LHC).

\vspace{-0.4cm}
\section{Results and Discussion}

Let us present the predictions for hard diffractive production of
W's based on the previous discussion. They are compared
with experimental data from \cite{CDF,D0} in Table I, where estimates for the LHC are also presented. In the numerical calculations, we have used the new H1 parameterizations for the diffractive pdf's. \cite{H1diff}  As the larger uncertainty comes from the gap survival factor, the error in the predictions correspond to the theoretical band for $<\!|S|^2\!>$. In the theoretical expressions of previous section one computes only the interaction of pomerons (emmitted by protons) with antiprotons (protons in LHC case), that means events with rapidity gaps on the side from which antiprotons come from.  Disregarding the gap factor, the diffractive production rate is approximately 7 \% (using the cut $|\eta|<1$) being very large compared to the Tevatron data. When considering the gap survival probability correction, the values are in better agreement with data. When considering central $W$ boson fraction, $-1.1<\eta_e<1.1$ (cuts of CDF and D0 \cite{CDF,D0}), we obtain a diffractive rate of 0.67 \% using the KMR estimate for $<\!|S|^2\!>$, whereas it reaches 0.76\% for the GLM estimate. The average rate considering the theoretical band for the gap factor is then $R_W= 0.715\pm 0.045$ \%.  Considering the forward $W$ fraction, $1.5<|\eta_e|<2.5$ (D0 cut), one obtains $R_W=0.83$ \% for KMR and $R_W=2.58$ \% for GLM, with an averaged value of $R_W= 1.7\pm 0.875$ \%. In this case, our estimate is larger than the central experimental value $R_W^{\mathrm{D0}}=0.64$ \%. For the total $W\rightarrow e\nu$ we have $R_W=0.68$ \% for KMR and $R_W=0.79$ \% for GLM  and the mean value $R_W= 0.735 \pm 0.055$ \%, which is in agreement with data and consistent with a large forward contribution. Finally, we estimate the diffractive ratio for LHC energy, $\sqrt{s}=14$ TeV. In this case we extrapolate the pdf's in proton and diffractive pdf's in Pomeron to that kinematical region. This procedure introduces somewhat additional uncertainties in the theoretical predictions. We take the conservative cuts $|\eta_e|<1$, $E_{T_{\mathrm{min}}}=20$ GeV for the detected lepton and $x_{\pom}<0.1$. We find $R_W=32.7$ \% for KMR gap survival probability factor and $R_W=29.5$ \% for GLM, with a mean value of $R_W^{\mathrm{LHC}}=31.1\pm 1.6 $ \%. This means that the diffractive contribution reaches one third, or even more, of the inclusive hadroproduction even when multi-Pomeron scattering corrections are taken into account.

In summary, we have shown that it is possible to obtain a reasonable
overall description of hard diffractive hadroproduction of $W$ relying on Regge factorization including the suppression from  gap survival probability factor. We find that the ratio of diffractive to non-diffractive boson production is in good agreement with the CDF and D0 data when considering these corrections. The overall diffractive ratio for $\sqrt{s}=1.8$ TeV (Tevatron) is of order 1 \%. In addition, we make predictions which could be compared to future measurements at LHC. The estimates give large rates of diffractive events, reaching values higher than 30~\% of the inclusive cross section.
\vspace{-0.2cm}
\begin{table}[t]
\tbl{Data versus model predictions for diffractive $W^{\pm}$ hadroproduction (cuts $E_{T_{\mathrm{min}}}=20$ GeV and $x_{\pom}<0.1$). }
{\begin{tabular}{@{}cccc@{}} \toprule
$\sqrt{s}$  & Rapidity  & Data (\%) & Estimate (\%)\\
\hline
1.8 TeV &    $|\eta_e|<1.1$  & $1.15\pm 0.55$ \cite{CDF}  & $0.715\pm 0.045$\\
1.8 TeV &    $|\eta_e|<1.1$  & $1.08\pm 0.25$ \cite{D0} & $0.715\pm 0.045$\\
1.8 TeV &    $1.5<|\eta_e|<2.5$  & $0.64\pm 0.24$  \cite{D0} & $1.7\pm 0.875$\\
1.8 TeV &     Total $W\rightarrow e\nu $ & $0.89\pm 0.25$ \cite{D0} & $0.735\pm 0.055$ \\
14 TeV &    $|\eta_e|<1$  & ---  & $31.1\pm 1.6$\\
\botrule
\end{tabular}}
\end{table}


\vspace{-0.4cm}

\begin{thebibliography}{0}

\bibitem{Collins} P. D. B. Collins, {\it An Introduction to Regge theory
and high energy physics} (Cambridge University Press, Cambridge, England, 1977).

\bibitem{IS} G. Ingelman and P.E. Schlein, {\it Phys. Lett.}  {\bf B152}
(1985) 256.

\bibitem{H1diff} H1 Collab.,  A. Aktas {\it et al.}, {\it Eur. Phys. J.} {\bf C48} (2006) 715.

\bibitem{KKMR} A.B.~Kaidalov, V.A.~Khoze, A.D.~Martin and
M.G.~Ryskin, {\it Eur. Phys. J.} {\bf C21} (2001) 521.

\bibitem{ggm_prd} M.B. Gay Ducati, M.M. Machado and M.V.T. Machado, arXiv:hep-ph/0703315.

\bibitem{GLM} E.~Gotsman, E.~Levin and U.~Maor, {\it Phys. Rev.} {\bf
D60} (1999) 094011.

\bibitem{CDF} CDF Collab., F. Abe {\it et al.}, {\it Phys. Rev. Lett.}
{\bf 78} (1997) 2698.

\bibitem{D0} D0 Collab., V.M. Abazov {\it et al.}, {\it Phys. Lett.}
{\bf B574} (2003) 169.



\end{thebibliography}
\end{document}